\documentclass{ws-4-MG14}            
\begin{document}
\title{Buchert coarse-graining and the classical energy conditions}
\author{Matt Visser}
\address{School of Mathematics and Statistics, Victoria University of Wellington,\\
Wellington 6140, New Zealand\\
E-mail: matt.visser@msor.vuw.ac.nz\\
http://www.victoria.ac.nz/sms}
\begin{abstract}
So-called ``Buchert averaging'' is actually a coarse-graining procedure, where fine detail is ``smeared out'' due to limited spatio-temporal resolution. 
For technical reasons, (to be explained herein), ``averaging'' is not really an appropriate term, and I shall consistently describe the process as a  ``coarse-graining''.  
Because Einstein gravity is nonlinear  the  coarse-grained Einstein tensor is typically not equal to
the Einstein tensor of the coarse-grained spacetime geometry. 
The discrepancy can be viewed as an ``effective" stress-energy.
To keep otherwise messy technical issues firmly under control, I shall work with  conformal-FLRW (CFLRW) cosmologies. 
These CFLRW-based models are
particularly tractable, and are also particularly attractive observationally: the CMB is not distorted. 
In this CFLRW context  one can  prove some rigorous theorems regarding the interplay between 
Buchert coarse-graining, tracelessness of the effective stress-energy,  and the classical energy conditions.
\end{abstract}


\bodymatter
\def\d{{\mathrm{d}}}
\def\O{{\mathcal{O}}}
\section{Introduction}

The cosmological back-reaction problem continues to generate considerable (and sometimes quite heated) debate.~\cite{Green:2010, Green:2013, Green:2014, Green:2015,  Buchert:2011, Ellis:2011, Buchert:2011b, Wiltshire:2011, Buchert:2015, Buchert:2015b, Ostrowski:2015} 
There is significant disagreement as to just how one should split the universe into ``smooth background'' plus ``local perturbations'', and yet more disagreement as to whether or not these perturbations remain small. 
I shall work within the framework of conformally-FLRW (CFLRW) cosmologies,~\cite{Visser:2015} where the mathematics and physics are both firmly under control.~\cite{Visser:2015} 
These CFLRW cosmologies are simply generic FLRW spacetimes distorted by a (possibly non-perturbatively large) conformal factor. 
Cosmographically~\cite{Visser:2004} this is the unique non-perturbative distortion that can be applied to FLRW cosmologies without grossly modifying the CMB.~\cite{Visser:2015} 
In counterpoint, when working in this CFLRW framework, Buchert's coarse-graining procedure simplifies tremendously. 
The contribution to the effective stress-energy due to coarse graining can be explicitly calculated. 
This coarse-graining contribution to the effective stress-energy need not be traceless, though it will typically satisfy the classical energy conditions. 
(Further details are in preparation.~\cite{Visser:2015b})

\section{Strategy}
Let $g_{ab}$ represent a FLRW spacetime.
Adopt conformal time for the FLRW. 
Then:
\begin{equation}
g_{ab} \; \d x^a \; \d x^b  = a(\eta)^2 \left\{ -\d \eta^2 + {\d r^2\over 1-kr^2} + r^2 \d \Omega^2\right\} 
= a(\eta)^2 \;\{\hat g_{ab} \; \d x^a \; \d x^b \}.
\end{equation}
Here $\hat g_{ab}$ is translation invariant in both space and (conformal) time.
On this FLRW space-time, define coarse-graining by
\begin{equation}
\langle  \phi(x) \rangle = \oint f(x-y) \; \phi(y) \; \d^4 y; \qquad  \oint f(x-y) \; \d^4 y = 1.
\end{equation}
Here $f(x-y)$ is any non-negative translation-invariant normalized kernel in both space and (conformal) time.
The high symmetries of FLRW are essential to this construction.

\section{Preliminaries}

In this FLRW framework, coarse-graining the gradient of a scalar is the same as taking the gradient of the coarse-grained  scalar:
\begin{eqnarray}
\langle \nabla_x \phi(x) \rangle &=& \oint f(x-y) \; \nabla_y \phi(y) \;\d^4 y 
\\
&=& - \oint   [\nabla_y f(x-y)] \; \phi(y) \;\d^4 y 
\\
&=&  + \oint   [\nabla_x f(x-y)]\; \phi(y) \; \d^4 y
\\
&=&  \nabla_x  \oint   f(x-y)\; \phi(y) \;\d^4 y 
\\
&=& \nabla_x \langle \phi(x) \rangle.
\end{eqnarray}
Translation invariance of the kernel is crucial to establishing this result.

Now calculate:
\begin{eqnarray}
\langle \langle \phi(x) \rangle \rangle &=& \oint f(x-y) \;  \langle \phi(y) \rangle \;\d^4 y 
\\
&=& \oint f(x-y) \; \left( \oint f(y-z) \;  \phi(z) \;\d^4 z \right)  \;\d^4 y 
\\
&=&  \oint \left(\oint f(x-y) \;   f(y-z) \;  \d^4 y \right) \phi(z) \;\d^4 z
\\
&=&  \oint \left(\oint f(x-z) \;   f(z-y) \;  \d^4 z \right) \phi(y) \;\d^4 y
\\
&=& \oint \left[f^{\otimes2}\right](x-y) \;  \; \phi(y) \;\d^4 y
\\
&\neq&  \langle \phi(x) \rangle
\end{eqnarray}
So you have to be very careful. Since $\langle \langle \phi(x) \rangle \rangle  \neq  \langle \phi(x) \rangle$, then if you insist on using the word ``averaging'' to describe the coarse-graining process, it follows that ``the average of the average is not the average''. This is at gross variance with the normal use of the word ``averaging'', and for this reason I shall eschew this term. 
Furthermore:
\begin{equation}
\left\langle \left(\vphantom{\big|}\phi(x) - \langle \phi(x) \rangle\right)^2 \right\rangle  \neq  \langle \phi(x)^2 \rangle -  \langle \phi(x) \rangle^2.
\end{equation}
More precisely
\begin{eqnarray}
\left\langle \left(\vphantom{\big|}\phi(x) - \langle \phi(x) \rangle\right)^2 \right\rangle  &=&  \langle \phi(x)^2 \rangle -  \langle \phi(x) \rangle^2
\nonumber\\
&&
+ \langle \langle \phi(x) \rangle^2 \rangle  - 2 \langle \phi(x) \langle \phi(x) \rangle \rangle +  \langle \phi(x) \rangle^2.
\end{eqnarray}
For ensemble averages, or indeed any situation where the normal rules of averaging apply, that second line vanishes.
But it does {not} vanish for a coarse-graining process.
For these reasons  the phrase ``coarse-graining'' is more appropriate than the word ``{averaging}''.
While  one cannot now naively assert  $\langle \phi(x)^2 \rangle -  \langle \phi(x) \rangle^2 \geq 0$ based on the usual ``averaging'' arguments, this inequality  can instead be derived from positivity of the coarse-graining kernel plus an appeal to the Cauchy--Schwarz inequality:
\begin{equation}
 \langle \phi(x) \rangle =  \int f(x-y) \phi(y) \d y \leq \sqrt{ \int f(x-y) \phi(y)^2 \d y \int f(x-y) \d y} = \sqrt{ \langle \phi(x)^2 \rangle }.
\end{equation}

\section{Conformally related spacetimes}

Consider first a conformal deformation of a given metric: $e^{2\phi}\; g_{ab}$.
We first note a number of  utterly standard results.
\begin{itemlist}
\item 
{Ricci tensor}:
\begin{equation}
R(e^{2\phi}g)_{ab} = R(g)_{ab} - 2 [\nabla_a\nabla_b \phi - \nabla_a \phi \nabla_b \phi] -  g_{ab} [\nabla^2\phi+2|\nabla\phi|^2].
\end{equation}
\item
{Ricci scalar}:
\begin{equation}
g^{ab} R(e^{2\phi}g)_{ab} = g^{ab} R(g)_{ab} -6 [\nabla^2\phi+|\nabla\phi|^2].
\end{equation}
\item
{Einstein tensor}:
\begin{equation}
G(e^{2\phi}g)_{ab} = G(g)_{ab} - 2 [\nabla_a\nabla_b \phi - \nabla_a \phi \nabla_b \phi] + g_{ab} [2\nabla^2\phi+|\nabla\phi|^2].
\end{equation}
\end{itemlist}
Now apply (generic) coarse-graining.
\begin{itemlist}
\item 
Coarse-grain the Einstein tensor:
\begin{equation}
\langle G(e^{2\phi}g)_{ab} \rangle= G(g)_{ab} 
- 2 \langle\nabla_a\nabla_b \phi - \nabla_a \phi \nabla_b \phi\rangle 
+ g_{ab} [\langle2\nabla^2\phi+|\nabla\phi|^2\rangle].
\end{equation}
\item
Coarse-grain the conformal mode:
\begin{equation}
G(e^{2\langle \phi\rangle}g)_{ab}= G(g)_{ab} 
- 2 \nabla_a\nabla_b \langle\phi\rangle - \nabla_a \langle\phi\rangle \nabla_b \langle\phi\rangle 
+ g_{ab}[ 2\nabla^2\langle\phi\rangle+|\nabla\langle\phi\rangle|^2].
\end{equation}
\item
Coarse-grain  the conformally deformed metric holding the base metric fixed:
\begin{eqnarray}
G(\langle e^{2\phi}g\rangle)_{ab}= G(\langle e^{2\phi}\rangle g)_{ab}&=& G(g)_{ab} 
- \left[\nabla_a\nabla_b \ln\langle e^{2\phi}\rangle 
- {1\over2} \nabla_a \ln\langle e^{2\phi}\rangle \nabla_b \ln\langle e^{2\phi}\rangle \right]
\nonumber
\\
&& +  g_{ab} \left[\nabla^2\ln\langle e^{2\phi}\rangle+{1\over4}|\nabla\ln\langle e^{2\phi}\rangle|^2\right].
\end{eqnarray}
\end{itemlist}
This so far holds for any generic coarse-graining process, details as yet unspecified. (We have not even used any assumed translation invariance for the kernel in the coarse-graining process.)

\section{CFLRW spacetimes}

For CFLRW space-time smearing this simplifies considerably.
\begin{itemlist}
\item 
Coarse-grain the Einstein tensor:
\begin{equation}
\langle G(e^{2\phi}g)_{ab} \rangle= G(g)_{ab} 
- 2 \nabla_a\nabla_b \langle \phi \rangle - \langle\nabla_a \phi \nabla_b \phi\rangle 
+ g_{ab} [2\nabla^2\langle\phi\rangle+\langle|\nabla\phi|^2\rangle].
\end{equation}
\item
Coarse-grain the conformal mode:
\begin{equation}
G(e^{2\langle \phi\rangle}g)_{ab}= G(g)_{ab} 
- 2 \nabla_a\nabla_b \langle\phi\rangle - \nabla_a \langle\phi\rangle \nabla_b \langle\phi\rangle 
+ g_{ab}[ 2\nabla^2\langle\phi\rangle+|\nabla\langle\phi\rangle|^2].
\end{equation}
\item
Coarse-grain the metric:
\begin{eqnarray}
G(\langle e^{2\phi}g\rangle)_{ab}= G(\langle e^{2\phi}\rangle g)_{ab}&=& G(g)_{ab} 
- \left[\nabla_a\nabla_b \ln\langle e^{2\phi}\rangle 
- {1\over2} \nabla_a \ln\langle e^{2\phi}\rangle \nabla_b \ln\langle e^{2\phi}\rangle \right]
\nonumber
\\
&&
+  g_{ab} \left[\nabla^2\ln\langle e^{2\phi}\rangle+{1\over4}|\nabla\ln\langle e^{2\phi}\rangle|^2\right].
\end{eqnarray}
\end{itemlist}
Combining the above, for CFLRW coarse-graining we have:
\begin{eqnarray}
\langle G(e^{2\phi}g)_{ab} \rangle &=& G(e^{2\langle \phi\rangle}g)_{ab}
 - 2[ \langle \nabla_a \phi \nabla_b \phi\rangle  -\nabla_a \langle\phi\rangle \nabla_b \langle\phi\rangle ]
 + g_{ab} [\langle|\nabla\phi|^2\rangle - |\nabla\langle\phi\rangle|^2].
 \nonumber
 \\
 &&
\end{eqnarray}
This is a specific, (very high symmetry), case of Buchert coarse-graining, 
one where all the technical details are now fully under control.

\section{Effective stress-energy}

The coarse-grained spacetime ``sees'' the effective stress-energy
\begin{equation}
T_{ab}^\mathrm{effective} = \langle T_{ab} \rangle + \Delta T_{ab}^\mathrm{Buchert}.
\end{equation}
The part of the ``effective stress tensor'' due to CFLRW coarse-graining is:
\begin{equation}
\Delta T^\mathrm{Buchert}_{ab} =   2[ \langle \nabla_a \phi \nabla_b \phi\rangle  -\nabla_a \langle\phi\rangle \nabla_b \langle\phi\rangle ] - g_{ab} [\langle|\nabla\phi|^2\rangle - |\nabla\langle\phi\rangle|^2].
\end{equation}
Note that this vanishes (as it should) in the limit where the smearing kernel becomes a delta function (so that coarse-graining is switched off). 
For the trace of this Buchert contribution to the effective stress tensor we have
\begin{equation}
\Delta T^\mathrm{Buchert} =   -2 \left\{\langle|\nabla\phi|^2\rangle - |\nabla\langle\phi\rangle|^2\right\} =
 2 \left\{\langle \dot\phi^2\rangle - \nabla\dot\phi\rangle^2\right\}
 -2 \sum_{i=1}^3\left\{\langle\nabla_i\phi^2\rangle - \nabla_i\langle\phi\rangle^2\right\}.
\end{equation}
This need not be zero, and generically is in fact nonzero. Since the first term is positive and the last three terms are negative, the trace can easily be arranged to be either positive or negative. 
In contrast, this effective stress energy typically will satisfy many of the other classical {energy conditions}.
\enlargethispage{20pt}

\noindent
--- For instance, for null vectors, using the FLRW symmetries to assert $k^a \nabla_a \langle\phi\rangle= k^a\langle  \nabla_a\phi\rangle= \langle k^a \nabla_a \phi\rangle$, we have:~\cite{Visser:1997, Visser:1997b, Visser:1997c, Visser:1999, Barcelo:2002, Visser:1994}
\begin{equation}
\Delta T^\mathrm{Buchert}_{ab} \;k^a \;k^b=    2 \left\{ \langle (k^a \nabla_a \phi)^2\rangle -  \langle k^a \nabla_a \langle\phi\rangle^2\right\} \geq 0.
\end{equation}
So the null energy condition (NEC) is  satisfied for this form of coarse-graining.

\noindent
--- For timelike vectors, we have:~\cite{Visser:1997, Visser:1997b, Visser:1997c, Visser:1999, Barcelo:2002, Visser:1994}
\begin{equation}
\Delta T^\mathrm{Buchert}_{ab} \;V^a \;V^b=    \left\{g^{ab}+2V^aV^b\right\} \;
\left\{ \langle \nabla_a \phi \nabla_b \phi \rangle -  \nabla_a \langle\phi\rangle \nabla_b \langle\phi\rangle \right\}.
\end{equation}
Use a tetrad decomposition with $e_0{}^a = V^a$ being the timelike vector $V^a$ of interest, then
\begin{equation}
g^{ab}+2V^aV^b = V^a V^b +\sum_{A=1}^3 e_A{}^a \, e_A{}^b  =  \sum_{A=0}^3 e_A{}^a \, e_A{}^b.
\end{equation}
This is a positive definite Euclidean signature ``metric''. Now using the FLRW symmetries to assert $e_A{}^a \nabla_a \langle\phi\rangle= e_A{}^a\langle  \nabla_a\phi\rangle= \langle e_A{}^a \nabla_a \phi\rangle = \langle \nabla_{e_A} \phi\rangle$,  we have
\begin{equation}
\Delta T^\mathrm{Buchert}_{ab} \;V^a \;V^b=   \sum_{A=0}^3  
\left\{\langle(e_A{}^a  \nabla_a \phi)^2\rangle -  \langle e_A{}^a\nabla_a\phi\rangle^2  \right\} \geq 0.
\end{equation}
So the weak energy condition (WEC) is  satisfied for this form of coarse-graining.

\noindent
--- For the strong energy condition (SEC) consider the trace-reversed quantity~\cite{Visser:1997, Visser:1997b, Visser:1997c, Visser:1999, Barcelo:2002, Visser:1994}
\begin{eqnarray}
\overline{\Delta T^\mathrm{Buchert}_{ab}} &=& 
\Delta T^\mathrm{Buchert}_{ab} - {1\over2} g_{ab} (\Delta T^\mathrm{Buchert}_{cd} \, g^{cd}) 
\\
&=&   2[ \langle \nabla_a \phi \nabla_b \phi\rangle  -\nabla_a \langle\phi\rangle \nabla_b \langle\phi\rangle ].
\end{eqnarray}
Then for timelike vectors
\begin{eqnarray}
\overline{\Delta T^\mathrm{Buchert}_{ab}} \;V^a \;V^b &=&    
2\;
\left\{ \langle \nabla_a \phi \nabla_b \phi \rangle -  \nabla_a \langle\phi\rangle \nabla_b \langle\phi\rangle \right\} \; V^aV^b
\\
&=& 2\left\{ \langle (\nabla_V \phi)^2  \rangle -  \langle \nabla_V\phi\rangle^2\right\} \geq 0.
\end{eqnarray}
So the strong energy condition (SEC) is  satisfied for this form of coarse-graining.

\section{Cosmological setting --- $w$ parameter}

In a cosmological setting, on large enough regions, one might (approximately) hope:
\begin{equation}
\langle \nabla_i \phi \nabla_j \phi\rangle =  \langle(\phi')^2\rangle\; g_{ij};
\qquad \langle \dot\phi\; \nabla_i \phi \rangle =0;  
\qquad \langle \nabla_i  \phi\rangle = 0.
\end{equation}
Then
\begin{equation}
\langle \nabla_a \phi \nabla_b \phi\rangle \to 
\left[ \begin{array}{c|c}
\vphantom{\Big|} \langle(\dot\phi)^2\rangle & 0 \\
\hline
0 &\vphantom{\Big|} \langle(\phi')^2\rangle \;g_{ij}
\end{array}\right],
\end{equation}
implying
\begin{equation}
\Delta T^\mathrm{Buchert}_{ab} \to \left[ \begin{array}{c|c}
\vphantom{\Big|} \langle(\dot\phi)^2\rangle +3 \langle(\phi')^2\rangle  - \langle\dot\phi\rangle^2& 0 \\
\hline
0 &\vphantom{\Big|} \left(\langle(\dot\phi)^2\rangle - \langle(\phi')^2\rangle -  \langle\dot\phi\rangle^2\right)g_{ij}
\end{array}\right].
\end{equation}
Thus
\begin{equation}
\Delta (\rho+3p)^\mathrm{Buchert} \to 4 \left(\langle(\dot\phi)^2\rangle  -  \langle\dot\phi\rangle^2\right) \geq 0.
\end{equation}
This agrees with the SEC calculation.
The ``effective $w$ parameter'' is
\begin{equation}
w =  {\langle(\dot\phi)^2\rangle - \langle(\phi')^2\rangle -  \langle\dot\phi\rangle^2 
\over 
 \langle(\dot\phi)^2\rangle +3 \langle(\phi')^2\rangle  - \langle\dot\phi\rangle^2}
\end{equation}
This contribution to stress-energy always has $w \in(-1/3,+1)$.

\section{Discussion}

When studying Buchert coarse-graining, working with CFLRW spacetimes is a great technical help.
Specifically, a FLRW background provides you with a very nicely controlled version of Buchert coarse-graining, 
and CFLRW coarse-graining satisfies some nice differential identities.
Indeed CFLRW coarse-graining yields an explicit and tractable effective stress-energy.
The effective stress energy is not necessarily traceless.
Furthermore for this form of coarse-graining the classical energy conditions {are satisfied} by the effective stress-energy.
Finally I emphasize that Buchert coarse-graining {$\neq$} ensemble average; several key properties are radically different.


\enlargethispage{10pt}

\end{document}